\documentstyle[aps,pra,multicol,amssymb,amsfonts]{revtex}

\newcommand{\abs}[1]{|#1|}
\newcommand{\norm}[1]{\| #1\|}
\newcommand{\trnorm}[1]{\| #1 \| _{T}}
\newcommand{\alg}[1]{\mbox{$\mathfrak{#1}$}}
\newcommand{\trace}[1]{\mbox{$\mathrm{Tr}$}(#1)}
\newcommand{\hil}[1]{\mbox{$\mathcal{#1}$}}
\newcommand{\dimn}{\mbox{$\mathrm{dim}$}}

\begin{document}
\title{Bipartite Mixed States of Infinite-Dimensional Systems
are Generically Nonseparable}

\author{Rob Clifton}
\address{Departments of Philosophy and History and Philosophy of
Science, \\ 10th floor, Cathedral of
Learning, University of
Pittsburgh, \\ Pittsburgh, PA\ 15260, USA. \\ email: rclifton+@pitt.edu}
\author{Hans Halvorson}
\address{ Department of Mathematics, 301 Thackeray Hall \\ 
\vskip 5pt
and
\vskip 5pt
Department of Philosophy, 1001 Cathedral of Learning \\ University of
Pittsburgh, Pittsburgh, PA\ 15260, USA. \\ email: hphst1+@pitt.edu }

\date{\today}
\maketitle
\begin{abstract}
  Given a bipartite quantum system represented by a Hilbert space
  \mbox{$\hil{H}_{1}\otimes \hil{H}_{2}$}, we give an elementary
  argument to show that if either $\dimn \hil{H} _{1}=\infty$
  \emph{or} $\dimn \hil{H}_{2}=\infty$, then the set of nonseparable
  density operators on $\hil{H}_{1}\otimes \hil{H}_{2}$ is trace-norm
  \emph{dense} in the set of all density operators (and the separable
  density operators \emph{nowhere} dense).  This result complements
  recent detailed investigations of separability, which show that when
  $\dim \hil{H}_{i}<\infty$ for $i=1,2$, there is a separable
  neighborhood (perhaps very small for large dimensions) of the
  maximally mixed state.  
  \end{abstract}
\draft
\begin{multicols}{2}

\section{Introduction}
One key feature that distinguishes classical from quantum information
theory is the availability, in quantum theory, of composite systems
with entangled states.  Often a composite system will itself form part
of a larger system in a pure state wherein that composite system is
entangled, and perhaps interacting in a noisy way, with some
`environment'.  In that case, the state of the composite system will
necessarily be mixed.  Unfortunately, the theoretical analysis of
mixed entangled states --- usually called nonseparable states --- is
somewhat more complicated than it is for the case of pure states.
This is so even in the simplest case of a bipartite system represented
by a tensor product of just two Hilbert spaces $\hil{H}_{1}\otimes
\hil{H}_{2}$ with dimensions $d_{1},d_{2}>1$.
  
For example, while there is a uniquely natural measure of entanglement
for pure states, based on von Neumann entropy\cite{naturalmeasure},
there are a number of apparently distinct yet useful measures for
nonseparable states \cite{nonaturalmeasure}.  Moreover, there are
nonseparable states that are `bound entangled', i.e., no pure
entanglement (in the form of singlet states) can be distilled from
them\cite{boundentangled}.  At a more fundamental level, while every
pure entangled state dictates nonlocal Bell correlations, i.e.,
violates some Bell inequality \cite{Bell}, nonseparable states need
not \cite{wern,pop}. Bell-correlated states must always be
nonseparable.  However, there is no known characterization of the
Bell-correlated states amongst those which are nonseparable, at least
beyond the simplest $d_{1}=d_{2}=2$ case \cite{noBell}. There is not
even an effective characterization of the nonseparable states
themselves, amongst the set of all mixed states (but see
\cite{horo2}). There is, however, a useful general \emph{necessary}
condition for separability, in terms of positivity of the partial
transposition of the state's density matrix\cite{peres} (and in the
cases $d_{1}=2,d_{2}=2\ \mbox{or}\ 3$, this condition is sufficient
for separability as well \cite{horo2}).

There is a further curious disanalogy between pure entangled and
(mixed) nonseparable states. While the former are always norm-dense in
the unit sphere of $\hil{H}_{1}\otimes \hil{H}_{2}$, the nonseparable
states are \emph{not} dense in the set of all mixed states when
$d_{1},d_{2}<\infty$.  This was brought to light recently in an
investigation of whether the very noisy mixed states exploited by
certain models of NMR quantum computing are truly
nonseparable\cite{NMR}.  Moreover, determining just how generic
nonseparable states are is motivated by the more fundamental question,
``Is the world \emph{more classical or more quantum}?''~\cite{horo1}.

More specifically, in the finite-dimensional case --- either with
$d_{1},d_{2}<\infty$, or with an arbitrary finite number $N$ of
two-dimensional Hilbert spaces which can be grouped together to form a
tensor product $\hil{H}_{1}\otimes \hil{H}_{2}$ with
$d_{1},d_{2}=\mbox{even}$ --- it turns out that there is always an
open neighborhood (or a set of nonvanishing measure) of separable
states~\cite{NMR,horo1,horo2,cave99}.  In particular, in such a case
there is a natural maximally mixed state $\frac{1}{n}I$ (where
$n=d_{1}d_{2}$), and there is a separable neighborhood of
$\frac{1}{n}I$.  It has been shown, however, that in the case of $N$
qubits, the size of this neighborhood decreases with increasing $N$
\cite{NMR,cave99}.  And in the case of two (arbitrary)
finite-dimensional Hilbert spaces, numerical evidence has been
obtained indicating a similar shrinking of the separable
neighborhood~\cite{horo1,horo3}.  These results have prompted the
question, ``Does the volume of the set of separable states really go
to zero as the dimension of the composite system grows, and how
fast?'' \cite{horo1}.

In particular, one might conjecture that \emph{at} the limit, where
$d_{1}=d_{2}=\infty$, the separable states should be ``negligible.''
Our objective is to confirm this particular conjecture.  In fact, we
shall show that when $d_{1}=\infty$ \emph{or} $d_{2}=\infty$, the set
of separable states is \emph{nowhere dense} (relative to the
trace-norm topology).

\section{Density of Nonseparable States in the Infinite Case}
Let $\alg{B}(\hil{H}_{1}\otimes \hil{H}_{2})$ denote the set of all
(bounded) operators on $\hil{H}_{1}\otimes \hil{H}_{2}$, and let
$\alg{T}\equiv\alg{T}(\hil{H}_{1}\otimes \hil{H}_{2})$ be the subset
of (positive, trace-$1$ operators) density operators.  Recall that
$\alg{T}$ is a convex set; that is, for any $\{ D_{i}:i=1,\dots, n
\}\subseteq \alg{T}$, and for any sequence $\{ \lambda _{i} \}$ of
positive real numbers summing to $1$, $\sum _{i=1}^{n}\lambda
_{i}D_{i}$ is also in $\alg{T}$.  Throughout, we shall consider
$\alg{T}$ as endowed with the trace-norm topology, defined by
$\trnorm{A}\equiv\trace{(A^{*}A)^{1/2}}$ (reserving the notation
`$\|A\|$' for the standard operator norm).  For $D\in \alg{T}$, $D$ is
said to be a \emph{product state} just in case there is a $D_{1}\in
\alg{T}(\hil{H}_{1})$ and a $D_{2}\in \alg{T}(\hil{H}_{2})$ such that
$D=D_{1}\otimes D_{2}$.  The \emph{separable} density operators are
then defined to be all members of $\alg{T}$ that may be approximated
(in trace-norm) by convex combinations of product states~\cite{wern}.
In other words, the separable density operators are those in the
closed convex hull of the set of all product states in $\alg{T}$.

In what follows, we shall make use of a third, auxiliary Hilbert space
$\hil{H}_{3}$ with dimension $d_{3}=\infty$.  Let $\hil{S}$ denote the
closed unit sphere of $\hil{H}\equiv\hil{H}_{1}\otimes
\hil{H}_{2}\otimes \hil{H}_{3}$, endowed with the vector-norm
topology.  For $v\in \hil{S}$, let $\Phi (v)$ denote the unique
reduced density operator $D\in\alg{T}(\hil{H}_{1}\otimes \hil{H}_{2})$
such that $\langle v ,(A\otimes I)v\rangle =\trace{DA}$ for all $A\in
\alg{B}(\hil{H}_{1}\otimes \hil{H}_{2})$.  It is not difficult to see
that the reduction mapping $\Phi:\hil{S}\rightarrow\alg{T}$ is both
continuous and onto.

To see the continuity of $\Phi$, set $\abs{C}=(C^{*}C)^{1/2}$.
Using the polar decomposition of $C$, we have $\abs{C}=VC$, where $V$
is the partial isometry with initial space the closure of the range of
$C$ and final space the closure of the range of $C^{*}$ (cf. \cite[p. 4]{schat}).
If $C$ is trace-class, then \begin{equation}
  \trnorm{C} = \trace{\abs{C}} \: =\: |\trace{VC}| . \label{iso}
  \end{equation}  Now, for $u,v\in \hil{S}$, $\Phi (u)-\Phi (v)$ is
of trace-class.  Thus, using (\ref{iso}) with $C\equiv \Phi (u)-\Phi (v)$,
\begin{eqnarray}
\trnorm{\Phi (u) -\Phi (v) }&= &\Bigl| \trace{V\Phi (u) }
-\trace{V\Phi(v) } \Bigr| \\
&= & \Bigl| \langle u,(V\otimes I)u\rangle -\langle v,(V\otimes I)v\rangle \Bigr| \\
&= & \Bigl| \langle u,(V\otimes I)u\rangle - \langle u,(V\otimes I)v\rangle \nonumber
\\ & & +\langle u,(V\otimes I)v\rangle -\langle v,(V\otimes
I)v\rangle \Bigr| \\
&\leq &2\norm{u-v} ,\end{eqnarray} since $\norm{V\otimes I}=\norm{V}=1$ and
$u,v$ are unit vectors.  Thus, $\Phi:\hil{S}\rightarrow\alg{T}$ is
continuous.  It follows from this that $\Phi$ maps any dense set in
$\hil{S}$ onto a dense set in its image $\Phi( \hil{S})$.

To see that the mapping $\Phi$ is onto, let $D\in \alg{T}$.
Then, $D=\sum _{i=1}^{d_{1}d_{2}}\lambda
_{i}P_{x_{i}}$, where $\{ x_{i}:i=1,\dots ,d_{1}d_{2} \}$ is an orthonormal
family in
$\hil{H}_{1}\otimes \hil{H}_{2}$ (and $P_{x_{i}}$ projects onto the
ray $x_{i}$ generates).  Since $d_{3}=\infty$, there is always an orthonormal
family $\{ y _{i}:i=1,\dots d_{1}d_{2} \}$ in $\hil{H}_{3}$.  Thus
\begin{equation} v\:\equiv \:
\sum _{i=1}^{d_{1}d_{2}}\sqrt{\lambda _{i}}(x_{i}\otimes y_{i}) \end{equation}
defines a vector in $\hil{S}$ for which $\Phi (v)=D$.

We shall denote the norm-closure of any subset
$\hil{R}\subseteq\hil{H}$ by $[\hil{R}]$.  For any vector $v\in
\hil{H}$, we say that $v$ is $1$-\emph{cyclic} just in case the closed
subspace defined by
\begin{equation}
[(\alg{B}(\hil{H}_{1})\otimes I\otimes I)v]\equiv\Bigl[\{(A\otimes I\otimes
I)v :A\in \alg{B}(\hil{H}_{1})\}\Bigr]
\end{equation}
is the \emph{whole} of $\hil{H}=\hil{H}_{1}\otimes \hil{H}_{2}\otimes
\hil{H}_{3}$. With a $1$-cyclic state vector, one can get arbitrarily
close to any other vector in $\hil{H}$ simply by acting on the vector
with operators on the Hilbert space of system $1$.  Note that there is
no restriction to acting only with unitary operators or projections.
Nevertheless, states $v$ that are $1$-cyclic have an intuitive
physical interpretation.  Such states are all `maximally
EPR-correlated' in the following sense: For \emph{any} state vector
$w$ of system $2+3$ and \emph{any} $\epsilon>0$, there is a
measurement one can perform on system $1$ in state $v$ such that,
upon conditionalizing on an appropriate measurement outcome, the
probability that system $2+3$ is in state $w$ will be greater than
$1-\epsilon$ \cite{us}.  This is reminiscent of
Schr\"{o}dinger's\cite{scho} sardonic remark about the ``sinister''
possibility in quantum mechanics of steering a distant system (here,
system $2+3$) into any desired state by a suitable local measurement (on
system $1$).  Our next task is to show that $1$-cyclicity of a state
$v\in\hil{H}$ entails that it induces a nonseparable density operator
on the first two systems.

We begin with a few elementary observations. Let $\hil{H}'$ be any
Hilbert space and let $D\in \alg{T}(\hil{H}')$.  Then, for any $A\in
\alg{B}(\hil{H}')$, $ADA^{*}$ is a positive, trace-class operator.
We may then define the operator $D^{A}$ as follows: 
\begin{equation}
D^{A}\equiv \left\{ 
\begin{array}{cl}
\frac{ADA^{*}}{\trnorm{ADA^{*}}} &\qquad \mbox{when}\ 
\trnorm{ADA^{*}}\neq 0 \\
0 &\qquad \mbox{otherwise}.  \end{array} \right. \end{equation}
Thus, $D^{A}$ is either a density operator or the zero operator.  Suppose now
that $D$ is a convex combination, 
\begin{equation}
D=\sum _{i=1}^{n} \lambda _{i}D_{i}\,, \end{equation} 
where each $D_{i}\in \alg{T}(\hil{H}')$.  Then, \begin{equation} 
ADA^{*}=\sum _{i=1}^{n}\lambda _{i}(AD_{i}A^{*}) \end{equation} 
and \begin{eqnarray} \sum _{i=1}^{n} \lambda
  _{i}\trnorm{AD_{i}A^{*}}&=&\sum
  _{i=1}^{n} \lambda _{i}\trace{AD_{i}A^{*}} \\
  &=&\mbox{$\mathrm{Tr}$} \Bigl( \sum _{i=1}^{n}\lambda _{i}AD_{i}A^{*}\Bigr) \\
  &=&\trace{ADA^{*}}\:=\:\trnorm{ADA^{*}} .\end{eqnarray} Thus, if
$ADA^{*}\not=0$, and we let
\begin{equation} \lambda ^{A}_{i}\equiv \lambda _{i} 
\frac{\trnorm{AD_{i}A^{*}}}{\trnorm{ADA^{*}}} , \label{eq:hi} \end{equation}  
then $\sum _{i=1}^{n}\lambda ^{A}_{i}=1$ and 
\begin{equation} D^{A}\equiv \frac{ADA^{*}}{\trnorm{ADA^{*}}}=\sum
_{i=1}^{n}\lambda ^{A}_{i} D ^{A}_{i} . \label{eq:ho} \end{equation}

\textbf{Lemma 1}  If $v\in \hil{S}$ is $1$-cyclic, then $\Phi (v)$ is
nonseparable.

\emph{Proof:} Let $\hil{H}'\equiv \hil{H}_{1}\otimes \hil{H}_{2}$ and let
$A\in \alg{B}(\hil{H}_{1})$ be such that $\norm{(A\otimes I\otimes
  I)v}=1$.  A straightforward calculation---using the definition of a
reduced density operator---shows that \begin{equation} \Phi [(A\otimes
  I\otimes I)v]=(A\otimes I)\Phi (v)(A\otimes I)^{*}\in \alg{T} .\end{equation}
Suppose $\Phi (v)$ is separable.  (We show that $v$ cannot be
$1$-cyclic.)  Then, $\Phi (v)=\lim _{n}W_{n}$, where each $W_{n}\in
\alg{T}$ is a convex combination of product states.  But then,
\begin{eqnarray}
(A\otimes I)\Phi (v)(A\otimes I)^{*}&=&(A\otimes I)\lim
_{n}W_{n} (A\otimes I)^{*} \\
&=&\lim _{n}(A\otimes I)W_{n}(A\otimes I)^{*} \\
&=&\lim _{n}W^{A\otimes I}_{n}.\end{eqnarray}
The penultimate equality follows since multiplication by a fixed element in
$\alg{B}(\hil{H}')$ is trace-norm continuous~\cite[p. 39]{schat}.  The
final equality holds since \begin{eqnarray*} \lefteqn{\lim _{n}\trnorm{(A\otimes I)W_{n}(A\otimes
  I)^{*}} } \qquad \quad \\ 
&= & \trnorm{(A\otimes I)\Phi (v)(A\otimes I)^{*}}=1 .\end{eqnarray*}  
Now, for fixed $n$, \begin{equation}
W_{n}=\sum _{i=1}^{k}\lambda _{i}(D_{1i}\otimes D_{2i}) \end{equation}
and hence, from~(\ref{eq:ho}), \begin{eqnarray}
W_{n}^{A\otimes I} &=&\sum _{i=1}^{k}\lambda
^{A\otimes I}_{i}(D_{1i}\otimes D_{2i})^{A\otimes I} \\
&=&\sum _{i=1}^{k}\lambda ^{A\otimes I}_{i}(D^{A}_{1i}\otimes D_{2i}) .\end{eqnarray} 
Thus, the density operator $(A\otimes I)\Phi (v)(A\otimes I)^{*}$ is
again a limit of convex combinations of product states.  Hence $\Phi 
[(A\otimes I\otimes I)v]$ is separable for any $A\in \alg{B}(\hil{H}_{1})$ (where 
$\norm{(A\otimes I\otimes I)v}=1$).
  
Suppose, for reductio ad absurdum, that $v$ is also $1$-cyclic.  If we
let $\hil{M}$ denote the set of unit vectors in $\hil{H}$ of the form
$(A\otimes I\otimes I)v$, for some $A\in \alg{B}(\hil{H}_{1})$, then
$\hil{M}$ is dense in $\hil{S}$.  (If $A_{n}v\rightarrow w\in
\hil{S}$, then we may replace the sequence $\{ A_{n} \}$ with the
sequence $\{ A_{n}/\norm{A_{n}v} \}$.)  However, from the argument of
the previous paragraph, $\Phi (\hil{M})$ consists entirely of
separable density operators.  Since the separable density operators
are closed, the trace-norm closure $\Phi (\hil{M})^{-}$ only contains
separable states.  Finally, since $\Phi$ is both onto and continuous,
and $\hil{M}$ is dense, \begin{equation}
  \alg{T}=\Phi(\hil{S})=\Phi(\hil{M}^{-})\subseteq\Phi(\hil{M})^{-}\subseteq\alg{T}.
    \end{equation}
    Therefore every density operator in $\alg{T}$ must be separable,
    which is absurd.  It follows that, if $\Phi (v)$ is separable, $v$
    cannot be $1$-cyclic.  \emph{QED}
    
    We shall need only one further lemma, for which we will require
    the following definition.  A vector $v\in\hil{H}$ is called
    \emph{separating} for the subalgebra $I\otimes
    \alg{B}(\hil{H}_{2}\otimes \hil{H}_{3})$ if $Av=0$, with $A\in
    I\otimes \alg{B}(\hil{H}_{2}\otimes \hil{H}_{3})$, entails that
    $A=0$.  The physical interpretation of a separating state vector
    $v$ is simply that every possible outcome of every possible
    measurement on system $2+3$ has a nonzero probability of being
    found.
    
    \textbf{Lemma 2} If $d_{1}=\infty$, then the set of $1$-cyclic
    vectors is dense in $\hil{S}$.

\emph{Proof:} First observe that if $v$ is \emph{not} $1$-cyclic, then
it cannot be separating for
$I\otimes \alg{B}(\hil{H}_{2}\otimes
\hil{H}_{3})$.  Indeed, if we let $P$ denote the orthogonal
projection onto the closed subspace $[(\alg{B}(\hil{H}_{1})\otimes
I\otimes I)v]$, then it follows that $Pv=v$ and $P$ is in $I\otimes
\alg{B}(\hil{H}_{2}\otimes \hil{H}_{3})$.  (It is easy to see that $P$
must commute with all elements in $\alg{B}(\hil{H}_{1})\otimes
I\otimes I$, because they leave the range of $P$ invariant.)
Moreover, if $[(\alg{B}(\hil{H}_{1})\otimes I\otimes I)v]\neq \hil{H}$, then
$I-P\neq 0$ yet $(I-P)v=0$.  With this in mind, it is sufficient to
establish that the set of state vectors separating for $I\otimes
\alg{B}(\hil{H}_{2}\otimes \hil{H}_{3})$ is dense in $\hil{S}$.

Let $v\in \hil{S}$.  Since $d_{1}=d_{2}d_{3}=\infty$, $v$ has a
Schmidt decomposition
\begin{equation} v\:=\:\sum _{i=1}^{\infty}a_{i} (x_{i}\otimes y_{i}) ,
  \label{schmidt} \end{equation} where
$\{ x_{i}\}\subseteq\hil{H}_{1}$, $\{
  y_{i}\}\subseteq\hil{H}_{2}\otimes \hil{H}_{3}$, are orthonormal
  bases, and $\{ a_{i}\}$ is a sequence of coefficients (not
  necessarily all nonzero).  If $a_{i}\neq 0$ for all $i$, then it is
  clear that $v$ is separating for $I\otimes
  \alg{B}(\hil{H}_{2}\otimes \hil{H}_{3})$ (since the terms in the
  expansion of $(I\otimes A)v$, for any
  $A\in\alg{B}(\hil{H}_{2}\otimes \hil{H}_{3})$, will again be
  orthogonal).

On the other hand, if $a_{i}=0$ for at least one $i$, then consider
the new (normalized) state vector
\begin{equation} u\:=\:\frac{\sum _{i=1}^{\infty} b_{i}(x_{i}\otimes y_{i})
}{\sqrt{\sum
_{i=1}^{\infty}\abs{b_{i}}^{2}}} ,\end{equation}
where $b_{i}=a_{i}$ if $a_{i}\neq 0$, $b_{i} \neq 0$ if $a_{i}=0$, and the
sequence $\{ b_{i}
:a_{i}=0 \}$ is square-summable.  Each such $u$ is separating for $I\otimes
\alg{B}(\hil{H}_{2}\otimes \hil{H}_{3})$.
Moreover, we can make $u$ as close as we wish to $v$ by choosing the
coefficients $\{ b_{i}
:a_{i} =0 \}$ arbitrarily small.  Thus, the set of separating state vectors
for $I\otimes
\alg{B}(\hil{H}_{2}\otimes \hil{H}_{3})$ is dense in $\hil{S}$.  Since all
separating vectors for
$I\otimes \alg{B}(\hil{H}_{2}\otimes \hil{H}_{3})$ are $1$-cyclic, the
set of $1$-cyclic state vectors
is dense in $\hil{S}$.  \emph{QED} 

We turn, finally, to proving our main theorem:

\textbf{Theorem} \emph{If $\dimn \hil{H} _{1}=\infty$ or $\dimn
\hil{H} _{2}=\infty$, then the set of nonseparable density operators
on $\hil{H}_{1}\otimes \hil{H}_{2}$ is trace-norm dense, and the set
of separable density operators nowhere dense}.

\emph{Proof:} If $d_{2}=\infty$ then we may interchange the roles of
$\hil{H}_{1}$ and $\hil{H}_{2}$.  Thus, we may assume that
$d_{1}=\infty$.  Recall again that the continuous reduction map
$\Phi:\hil{S}\rightarrow\alg{T}(\hil{H}_{1}\otimes \hil{H}_{2})$ is
onto.  By Lemma $2$, the set of $1$-cyclic state vectors is dense in
$\hil{S}$.  Thus, $\{ \Phi (v) :v \;\mbox{is $1$-cyclic}\, \}$ is
dense in $\alg{T}$.  However, Lemma $1$ entails that each element of
the latter set is nonseparable.

By definition, the set of separable states in $\alg{T}$ is closed.
Since the set of nonseparable states is dense, the set of separable
states has empty interior.  Thus, the separable states are nowhere
dense in $\alg{T}$.  \emph{QED}

\section{Conclusion}
The fact that the states of an infinite-dimensional bipartite system
are generically nonseparable may or may not find any direct practical
application in quantum information theory.  However, from the point of
view of fundamental quantum physics, the implications appear profound.
For example, it is well-known that one cannot have a
finite-dimensional representation of the canonical (anti-)commutation
relations for a single degree of freedom.  It follows that the
position-momentum state of any pair of spin-less particles, or,
indeed, the position-momentum/spin state of a \emph{single} particle,
will be generically nonseparable (with similar conclusions applicable
in the field-theoretic case).

Moreover, it would be interesting to know, again from the point of
view of fundamental physics, whether Bell-correlated, and hence
nonlocal, states are \emph{also} generic in the infinite case.  This
conjecture is given some credence by the fact that the most widely studied
case of states which are nonseparable but violate no Bell
inequalities---so-called `Werner states' \cite{wern,pop}---all involve mixing
the maximally mixed state with some nonseparable state, yet there is
no (strictly) maximally mixed state in the infinite case.  Assuming no
obvious `no-Bell' neighborhood is forthcoming, one might first attempt
to prove the analogue of our main theorem for Bell states by
establishing that density operators on $\hil{H}_{1}\otimes
\hil{H}_{2}$ induced by $1$-cyclic states violate a Bell inequality
(noting that the Bell states, just like the nonseparable ones, form an
open set in $\alg{T}(\hil{H}_{1}\otimes \hil{H}_{2})$). And observe
that while $1$-cyclicity of a state $v$ implies that it induces a
nonseparable reduced density operator on $\hil{H}_{1}\otimes
\hil{H}_{2}$ (Lemma 1), $1$-cyclicity is strictly stronger than
nonseparability.  (To see this, choose some state vector $v\in
\hil{S}$ such that $\Phi (v)$ is any entangled pure state in
$\alg{T}(\hil{H}_{1}\otimes \hil{H}_{2})$.  Then $v$ reduces to a pure
state in $\alg{T}(\hil{H}_{3})$, is therefore not separating for
$I\otimes\alg{B}(\hil{H}_{2}\otimes \hil{H}_{3})$, and hence could not
be $1$-cyclic.)  Finally, in the special case where \emph{both}
$d_{1}=d_{2}=\infty (=d_{3})$, it was shown in \cite{us} that the set
of all states that are simultaneously $1$-, $2$-, and $3$-cyclic is
\emph{also} dense in the unit sphere of $\hil{H}=\hil{H}_{1}\otimes
\hil{H}_{2}\otimes\hil{H}_{3}$.  It would be surprising if even one
such `tricyclic' state were to induce a density operator on
$\hil{H}_{1}\otimes \hil{H}_{2}$ without Bell correlations.  Barring
surprises, our conjecture is, then, that our main theorem holds
(possibly only after replacing `or' by `and') for Bell-correlated
states as well.  If true, this would mean that there is essentially no
difference in the infinite case between Bell-correlated and
nonseparable states---a dramatic simplification over the finite case.    

It should be emphasized, however, that while nonseparable states need
not be Bell-correlated, this does not make them entirely devoid of
nonlocal properties.  As Popescu\cite{pop} has shown, a large class of
Werner states contain ``hidden locality,'' in that they violate
extended Bell inequalities that involve the performance of consecutive
measurements on the two subsystems.  Moreover, this violation becomes
maximal approaching the limit $d_{1}=d_{2}=\infty$.  Werner states
have also been said to contain ``active nonlocality'' at the level of
each single particle pair, since they can realize teleportation with a
fidelity better than any classical procedure \cite{tel}.  Similar
investigations have not been undertaken on infinite-dimensional
Werner-like states (should there be any).  However, it would now
appear that, at least with regards to infinite-dimensional systems,
the world is \emph{far} more quantum than classical.

\emph{Acknowledgments:} The authors would like to thank J.L. Bell
(Western Ontario) for preventing a mathematical mishap.

\end{multicols}
\end{document}